# Chemical Pressure Tuning of Magnetic Coupling in the La$_{1.4}$Sr$_{1.6}$Mn$_2$O$_7$ Bilayered Manganite

*Lorenzo Malavasi$^{a,*}$, Clemens Ritter$^b$, M. Cristina Mozzati$^c$, Carlo B. Azzoni$^c$, and Giorgio Flor$^a$*

$^a$ Dipartimento di Chimica Fisica "M. Rolla", Viale Taramelli 16, Pavia, Italy.
$^b$ Institute Laue-Langevin, Boite Postale 156, F-38042, Grenoble, France.
$^c$ CNISM- Unità di Pavia and Dipartimento di Fisica "A. Volta", Via Bassi 6, Pavia, Italy.

In this paper the effect of the chemical pressure induced by the Sr replacement for Ca in the La$_{1.4}$Sr$_{1.6}$Mn$_2$O$_7$ lattice is directly probed by means of neutron diffraction as a function of temperature. The results show a progressive change of the magnetic coupling from ferromagnetic to antiferromagnetic by increasing the Ca content,. indicating that the orbital order of the Mn ions is affected in the same way by the application of an external pressure and by the chemical pressure.

*Corresponding Author: Dr. Lorenzo Malavasi, Dipartimento di Chimica Fisica "M. Rolla", INSTM, Università di Pavia, V.le Taramelli 16, I-27100, Pavia, Italy. Tel: +39-(0)382-987921 - Fax: +39-(0)382-987575 - E-mail: lorenzo.malavasi@unipv.it

Colossal magnetoresistive materials still represent a very active subject for the basic and applied research in the field of solid state physics[1,2]. Besides the hole doped $AMnO_3$ perovskites compounds, where A=Rare Earth, strong interest has been more recently triggered off by the bilayered manganites which represent the *n*=2 members of the Ruddlesden-Popper series $A_{n+1}Mn_nO_{3n+1}$[3-5]. This structure corresponds to the staggering of 2 perovskites layers intercalated by a rock-salt block. The reduced dimensionality of the this system with respect to the 3D perovskite manganites is a interesting playground to study the effects of charge carrier confinement on the electronic and magnetic properties.

In the *n*=2 member $La_{2-2x}Sr_{1+2x}Mn_2O_7$ an extremely rich variety of magnetic phases has been found as a function of Sr-doping (*x*). One of the most interesting regions extends from *x*=0.3 to *x*=0.4 where a ferromagnetic and metallic (FMM) behaviour has been observed. For the *x*=0.30 member of the solid solution, *i.e.* $La_{1.4}Sr_{1.6}Mn_2O_7$, Kimura *et al.* showed the presence of a ferromagnetic order within the constituent bilayers which sets up around 100 K and a FM order between adjacent bilayers which progressively develops by lowering the temperature[6]. As the Sr-doping increases, the magnetic coupling within the bilayers changes from FM to AFM at about *x*~0.50 as a consequence of the change in the spin-direction within the *a-b* plane.

The role of cation substitution on the magnetic properties in these systems has been less extensively studied with respect to the perovskites oxides. Few works dealt with the Ca-doped series, for which, however, the only available magnetic results have been determined by means of susceptibility measurements[7] on the $La_{1.4}Sr_{1.6}Mn_2O_7$ and $La_{1.2}Sr_{1.8}Mn_2O_7$ compositions[8]. The observed trend in the Curie temperatures ($T_C$) revealed a progressive reduction along with the Ca-doping. However, no further insight into the magnetic structure evolution with the Ca-doping was presented.

The effect of Ca-doping on the magnetic structure of the bilayered manganites appears to be an interesting aspect that needs still to be clarified. In particular, the chemical pressure effect induced by the replacement of Sr with the isovalent Ca can reveal important information about the stability of the magnetic ground state in these systems and the comparison between these data and the variations induced by an hydrostatic external pressure[9] may shed light on the relative importance of these two effects.

In order to understand the role of the Ca-doping on the magnetic properties of bilayered manganites we carried out a neutron diffraction study of the optimally doped $La_{1.4}Sr_{1.6}Mn_2O_7$ composition where the Sr was progressively replaced by Ca up to 50% which is the cation solubility threshold for this Sr-doping ($x=0.3$)[10]. The study was completed by magnetic susceptibility measurements on the same samples.

$La_{1.4}(Sr_{1.6-y}Ca_y)Mn_2O_7$ samples with $y = 0$, 0.1, 0.2, 0.4, 0.6 and 0.8 were synthesized by solid state reaction starting from proper amounts of $La_2O_3$, $Mn_2O_3$, $SrCO_3$ and $CaCO_3$ (Aldrich >99.99%). Pellets were prepared from the thoroughly mixed powders and allowed to react first at 1273 K for 72 hours and after at 1573 K for other 72 hours. Phase purity was checked by means of X-ray diffraction (Bruker D8 Advance). All the samples patterns can be perfectly indexed according to a tetragonal unit cell belonging to the *I*4/*mmm* space group (no. 139). Cell volume variation is linear with the Ca-content thus suggesting the formation a homogeneous solid-solution. As usually found in the bilayered manganites, small traces of a perovskite impurity phase have been detected (below 5%, as estimated by the Rietveld refinement). This impurity phase was included in the Rietveld refinement. X-ray absorption spectroscopy at the Mn-edge confirmed the same Mn-valence for all the sample studied.

Figure 1 shows the field-cooling (FC) magnetic susceptibility measurements at 100 G for the samples considered in the present work. All the samples display a transition from a paramagnetic to a ferromagnetic state (PF transition). By increasing the Ca concentration the Curie temperature ($T_C$) and the value of the molar susceptibility ($\chi_{mol}$) are progressively reduced. The $T_C$ values – taken at the inflection point of the $\chi$ vs. $T$ curves – are reported in Table 1.

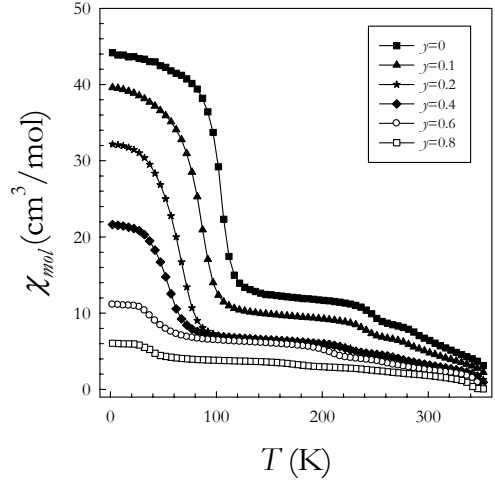

*Figure 1*: Field cooling measurements at 100 G for the $La_{1.4}(Sr_{1.6-yx}Ca_y)Mn_2O_7$ samples ($y$=0, 0.1, 0.2, 0.4, 0.6 and 0.8).

In addition, along with the Ca-substitution also the broadness of the magnetic transition increases, thus suggesting a progressive weakening of the ferromagnetic interactions between neighbouring Mn ions. However, the bulk magnetization measurements does not reveal the detail about the magnetic structure of the samples since it provides an information related to the overall magnetic response of the material. For this reason we further investigated this issue by means of neutron diffraction.

Figure 2 reports the neutron diffraction patterns for the $La_{1.4}(Sr,Ca)_{1.6}Mn_2O_7$ solid solution at 10 K, in the angular range where magnetic reflections appear. At this temperature the magnetic peaks (indicated by the arrows) are most intense for all the compositions.

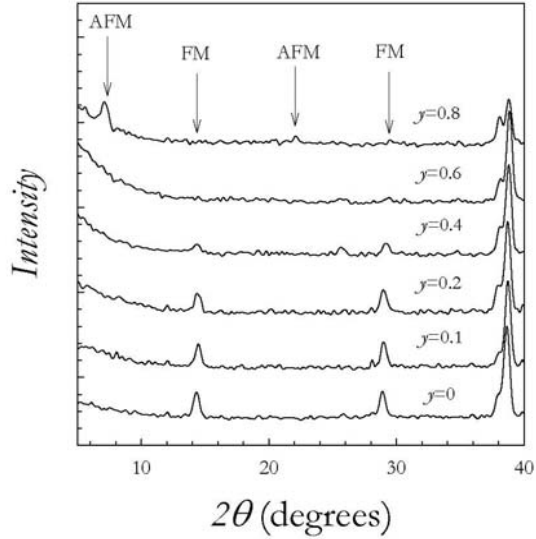

*Figure 2*: Neutron diffraction patterns for the $La_{1.4}(Sr_{1.6-y}Ca_y)Mn_2O_7$ samples ($y$=0, 0.1, 0.2, 0.4, 0.6 and 0.8) at the lowest investigated temperature, 10 K. Arrows mark some of the most intense magnetic reflections.

Through the refinement of the patterns as a function of temperature we observed that the magnetic structure of the Ca-undoped ($y$=0) compound can be very satisfactorily described by considering a magnetic cell where all the Mn spins are aligned parallel within each single bilayer and in all the successive bilayers. This leads to a FM coupling both in the single bilayers and between the bilayers. The spin direction has components both in the *a–b* plane and along the *c*-axis, forming an angle ($\theta$) of about 57°. This is consistent with the available literature data reporting the magnetic structure of Sr-doped bilayered manganites[4]. The Mn magnetic moment (in Bohr magnetons units) is reported in Figure 3. FM peaks start becoming evident at about 108 K (defined as *T**) and the value of the total Mn magnetic moment for the $La_{1.4}Sr_{1.6}Mn_2O_7$ sample at 10 K is 3.4(1) $\mu_B$, in good agreement with the calculated value of 3.70 $\mu_B$. We did not observe any appreciable variation of the canting angle $\theta$ with the temperature variation.

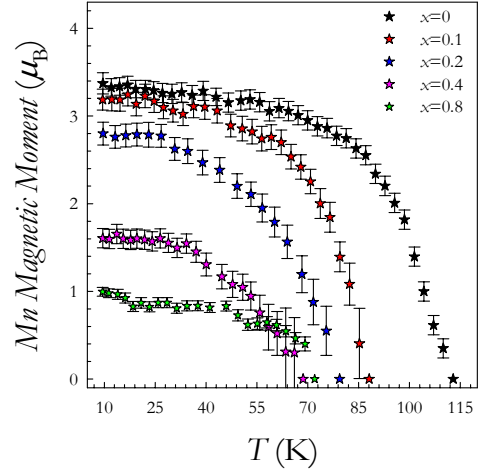

*Figure 3*: Refined magnetic moments for the La$_{1.4}$(Sr$_{1.6-y}$Ca$_y$)Mn$_2$O$_7$ samples (*y*=0, 0.1, 0.2, 0.4, 0.6 and 0.8) as a function of temperature.

By increasing the Ca-doping, thus at fixed hole doping, the magnetic structure found in the Ca-undoped sample is preserved until *y*=0.4. The relevant variations along with the cation replacement are the progressive reduction of the Mn magnetic moments (see Figure 3) and a reduction of *θ* which pass from 57° for *y*=0 to 46° for *y*=0.4. The values of the magnetic moments at 10 K and the temperatures at which the first magnetic peaks are observed (*T\**) are reported in Table 1.

By further increasing the Ca-doping to *y*=0.6 the neutron diffraction patterns, in all the *T*-range explored, do not reveal the presence of any magnetic reflection nor the intensity increase of any nuclear peaks even though this sample, as the other ones, displays a PF transition at about 37 K. A peculiar feature of the neutron diffractograms is the growth of a wide low-angle background below 10 ° which reaches its maximum in this sample. This can be related to the development of short-range magnetic interactions.

Finally, the further increase in the Ca concentration (*y*=0.8) leads to the appearance of magnetic reflections at about 75 K which are fully developed at 10 K, as can be appreciated by inspecting Figure 2. The magnetic peaks in this sample, however, are not compatible with a FM ordering. In contrast, the best agreement in the pattern refinement is achieved by considering an AFM ordering of A-type where each layer of the bilayer is FM ordered but the coupling between layers within the bilayer is AFM. Also in this case a component of the magnetic moment aligned along the *c*-axis is present. The refined total magnetic moment is listed in Table 1.

The overall data presented so far have shown clearly that the main effect of the Ca-substitution on the bulk properties of the $La_{1.4}(Sr_{1.6-y}Ca_y)Mn_2O_7$ solid solution (*y*=0, 0.1, 0.2, 0.4, 0.6 and 0.8) is a progressive reduction of the Curie temperature and of the molar susceptibility as *y* increases. However, by looking at the magnetic structure along the solid solution it is found that the Ca-substitution progressively weakens the FM long range magnetic order found at *y*=0 leading to its disappearance at about *y*=0.6 and finally to the development of an AFM structure at *y*=0.8. This behavior closely resembles the one found in the evolution of the magnetic structure in the $La_{2-2x}Sr_{2+2x}Mn_2O_7$ solid solution as a function of the Sr-doping: when changing the Sr concentration from *x*~0.32 to *x*~0.50 the magnetic coupling within the bilayer progressively changes from FM to AFM[4,6]. This behavior was linked to the progressive shift of electronic density from the $3z^2-r^2$ to the $x^2-y^2$ $e_g$ orbitals as the *x*-value increases.

A change of the magnetic coupling from FM to AFM has been reported in the previous literature for the $La_{1.4}Sr_{1.6}Mn_2O_7$ compound as a consequence of the application of an external pressure[9]. The Authors performed magnetization and neutron diffraction measurements as a function of the applied external pressure and concluded that the applied pressure induces the electron transfer from the out-of-plane to the in-plane orbital in Mn, thus enhancing the AFM order. This was also confirmed by a computational study which showed that the externally applied pressure controls the dimensionality of the orbital states and that a compression of the apical Mn-O bonds brings the bilayers closer to each other and stabilizes the in-plane orbital[11].

The results presented in this work strongly indicates that the chemical pressure effect caused by the Ca-doping mimics the influence of the hydrostatic pressure. The trend vs. $y$ of the normalized lattice parameters (Figure 4) clearly shows that the bilayered structure compresses strongly along the $c$-direction while slightly expands in the $a$-$b$ plane which agrees with a progressive electron transfer from $3d_{3z^2-r^2}$ to $3d_{x^2-y^2}$.

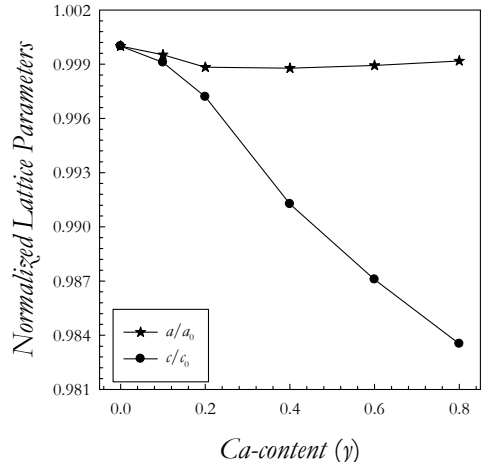

*Figure 4*: Trend of the normalized lattice parameters for the $La_{1.4}(Sr_{1.6-y}Ca_y)Mn_2O_7$ series as a function of the Ca-content, *y*.

In addition, we showed that the transition induced by the chemical pressure is a smooth one from a FM to a possible spin-glass state (around $y$=0.6) to a final AFM state. Let us remember that in the present case the Ca-doping does not induce any change in the hole doping in the material and, as a consequence, the effects reported are purely due to the structural changes caused by the cation substitution. Our data allows to conclude that, irrespective to the source of the compression of the bilayer, *i.e.* external or chemical pressure, the electronic structure rearranges in similar way thus stabilizing an in-plane occupancy of the Mn orbital which in turn leads to the stabilization of the AFM order. We remark that this paper provides the first direct evidence of the role of chemical pressure on the magnetic structure of $La_{1.4}Sr_{1.6}Mn_2O_7$ that until now had been only assessed by means of bulk magnetization measurements.

# Acknowledgments

ILL neutron facility and European Community financial support is acknowledged.

# Tables

| Ca content ($y$) | $T_C$ (K) | $T^*$ (K) | Total Magnetic Moment ($\mu_B$) at 10K |
|---|---|---|---|
| 0 | 102 | 108 | 3.4(1) |
| 0.1 | 82 | 82 | 3.2(1) |
| 0.2 | 67 | 68 | 2.8(1) |
| 0.4 | 52 | 51 | 1.6(1) |
| 0.6 | 37 | - | - |
| 0.8 | 37 | 75 | 1.0(1) |

*Table 1*: Curie temperatures ($T_C$), critical temperatures at which the first magnetic peaks are observed ($T^*$) and total magnetic moments at 10 K for the different Ca-doping.

# References

1. S. Jin, T. H. Tiefel, M. McCormack, R. A. Fastnacht, R. Ramesh, and L. H. Chen, Science **264**, 413 (1994).

2. J.M.D. Coey, M. Viret, and S. von Molnar, Advances in Physics **48**, 167 (1999).

3. T. Kimura, Y. Tomioka, H. Kuwahara, A. Asamitsu, M. Tamura, and Y. Tokura, Science, , **274**, 1698 (1996).

4. T. Kimura, and Y. Tokura, Annu. Rev. Mater. Sci, **30,** 451 (2000).

5. M.W. Kim, H.J. Lee, B.J. Yang, K.H. Kim, Y. Moritomo, J. Yu, and T.W. Noh, Phys. Rev. Lett., **98**, 187201 (2007).

6. T. Kimura, Y. Tomioka, A. Asamitsu, and Y. Tokura, Phys. Rev. Lett., **26**, 5920 (1998).

7. M. Takemoto, A. Katada, and H. Ikawa, Solid State Comm., **109**, 693 (1999).

8. C.H. Shen, R.S. Liu, S.F. Hu, J.G. Lin, C.Y. Huang, H.S. Sheu, J. Appl. Phys., **86**, 2178 (1999).

9. K.V. Kamenev, G.J. McIntyre, Z. Arnold, J. Kamarad, M.R. Lees, G. Balakrishnan, E.M.L. Chung, D.McK. Paul, Phys. Rev. Lett., **87**, 167203 (2001).

10. L. Malavasi, M.C. Mozzati, C. Tealdi, C.B. Azzoni, and G. Flor, Phys. Rev. B, **74**, 064104 (2006).

11. S. Ishihara, S. Okamoto, S. Maekawa, J. Phys. Soc. Jpn., **66**, 2965 (1997).

## Table of Contents Graphic

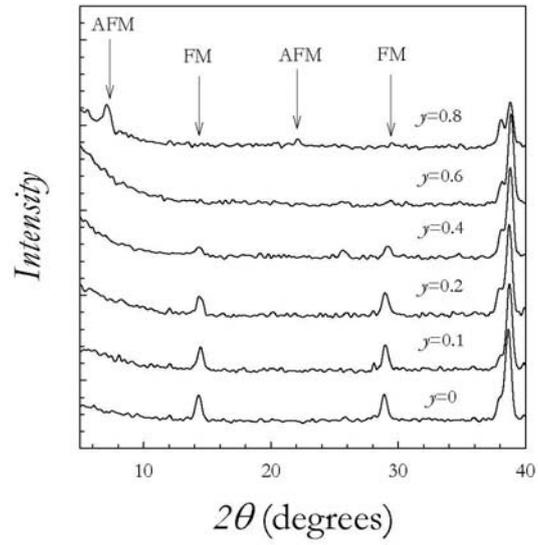

Evolution of the magnetic structure as a function of Ca doping ($y$) for the $La_{1.4}(Sr_{1.6-y}Ca_y)Mn_2O_7$ solid solution